\documentclass{emulateapj}
\usepackage{apjfonts}
\usepackage{natbib}
\usepackage{epsfig}

\newcommand{\etal}{et al.}

\newcommand{\Mdot}{\dot{M}}

\newcommand{\EV}[1]{\langle #1 \rangle}
\newcommand{\dlgL}{d\log(L)}
\newcommand{\eEdd}{l}
\newcommand{\nLP}{\dot{n}(L_{peak})}
\newcommand{\mdot}{\dot{m}}

\shorttitle{Quasar Lifetimes \&\ the Luminosity Function}
\shortauthors{Hopkins \etal}
\slugcomment{ApJ, Accepted, September 2005}

\begin{document}

\title{Luminosity-Dependent Quasar Lifetimes: \\
A New Interpretation of the Quasar Luminosity Function}
\author{Philip F. Hopkins\altaffilmark{1}, 
Lars Hernquist\altaffilmark{1}, 
Thomas J. Cox\altaffilmark{1}, 
Tiziana Di Matteo\altaffilmark{2}, 
%Paul Martini\altaffilmark{1}, 
Brant Robertson\altaffilmark{1}, 
Volker Springel\altaffilmark{3}}
\altaffiltext{1}{Harvard-Smithsonian Center for Astrophysics, 
60 Garden Street, Cambridge, MA 02138, USA}
\altaffiltext{2}{Carnegie Mellon University, 
Department of Physics, 5000 Forbes Ave., Pittsburgh, PA 15213}
\altaffiltext{3}{Max-Planck-Institut f\"{u}r Astrophysik, 
Karl-Schwarzchild-Stra\ss e 1, 85740 Garching bei M\"{u}nchen, Germany}

\begin{abstract}

We propose a new interpretation of the quasar luminosity function
(LF), derived from physically motivated models of quasar lifetimes and
light curves.  In our picture, quasars evolve rapidly and their
lifetime depends on both their instantaneous {\it and} peak
luminosities.  We study this model using simulations of galaxy mergers
that successfully reproduce a wide range of observed quasar phenomena.
With lifetimes inferred from the simulations, we deconvolve the
observed quasar LF from the distribution of peak luminosities, and
show that they differ qualitatively, unlike for the simple models of
quasar lifetimes used previously.  We find that the bright end of the
LF traces the intrinsic peak quasar activity, but that the faint end
consists of quasars which are either undergoing exponential growth to
much larger masses and higher luminosities, or are in sub-Eddington
quiescent states going into or coming out of a period of peak
activity.  The ``break'' in the LF corresponds directly to the maximum
in the intrinsic distribution of peak luminosities, which falls off at
both brighter and fainter luminosities.  Our interpretation of the
quasar LF provides a physical basis for the nature and slope of the
faint-end distribution, as well as the location of the break
luminosity.

\end{abstract}


\section{Introduction\label{sec:intro}}

The luminosity function (LF) of quasars is fundamental to cosmology,
but even after more than 30 years of study \citep[e.g.,][]
{Schmidt68,SG83,HS90,WHO94,Boyle00,Ueda03}, its
relation to the intrinsic luminosities of
quasars remains poorly understood.
Previous work modeling the quasar LF has relied on restrictive
assumptions about lifetimes and light curves of quasars, imagining,
for example, that quasars either have universal lifetimes or that they
evolve exponentially, usually on the galaxy dynamical time or the
$e$-folding time for Eddington-limited black hole growth $t_{S}=M_{\rm
BH}/\Mdot=4\times10^{8}\,\epsilon_{r}\,$yr for accretion with
radiative efficiency $\epsilon_{r}=L/\Mdot c^{2}\sim0.1$
\citep{Salpeter64}.  Under these circumstances, the distribution of
quasars with a given mass or peak luminosity is trivially related to
the observed LF (in the absence of selection effects), and the two
have essentially identical shape.

Recently, simulations of galaxy mergers incorporating black hole
growth and feedback \citep{SDH05b} have reproduced the $M_{BH}-\sigma$
relation between black hole mass and galaxy velocity dispersion
\citep{DSH05} and linked quasar activity \citep{H05a,H05b} to galaxy
evolution \citep{SDH05a}.  The simulations predict more complicated
quasar light curves than have been adopted previously.
The peak, exponential black hole growth phase is
determined by the gas supply over timescales $\sim 10^{8}\,{\rm yr}$
and shuts down when significant gas is expelled by feedback.  The
light curves have been studied by \citet{H05a,H05b}, who showed that
the self-termination process gives observable lifetimes
$\sim10^{7}\,$yr for bright optical quasars, in good agreement with
observations, and yields a large population of obscured sources as a
natural stage of quasar evolution. \citet{H05b} analyzed simulations
over a range of galaxy masses and found that the quasar light curves
and lifetimes are always qualitatively similar, with both the
intrinsic and observed quasar lifetimes being strongly decreasing
functions of luminosity, with longer lifetimes at all luminosities for
higher-mass (higher peak luminosity) systems.

Here, we use our quasar lifetimes and light curves to deconvolve the
observed quasar LF from the distribution of peak luminosities, and
find that they differ qualitatively, unlike for the trivial light
curves or even complex, cosmologically evolving light curves or 
distributions of Eddington ratios that have been employed previously, demanding a new
interpretation of the quasar LF.

\section{Modeling Quasar Lifetimes \&\ Light Curves\label{sec:lifetimes}}

Up to now, theoretical studies of the quasar LF have
generally employed very simple descriptions of the quasar light curve,
namely some variant of a ``feast or famine'' or ``light bulb'' model
\citep[e.g.,][]{SB92,KH00,WL03,HQB04}, in which quasars have only two
states: ``on'' or ``off.''  When ``on,'' quasars have a constant
luminosity and radiate at a characteristic Eddington ratio
$\eEdd\equiv\EV{L/L_{Edd}}$, generally in the range $\sim0.1-1$. This
state is assumed to last for a universal lifetime $t_{Q,\,LB}$,
which is the same for all quasars at a given redshift (although some
models allow for redshift evolution of $t_{Q,\,LB}$).  The value of
$t_{Q,\,LB}$ is an input parameter of the models, generally adopted
from observations or assumed to be related to the
Salpeter time or the dynamical time of the host galaxy.
The time spent in a logarithmic luminosity interval is then just
$dt/\dlgL = t_{Q,\,LB}\,\ln(10)\,L\,\delta(L-\eEdd\,L_{Edd})$.  Although this
approach is analytically simple, with the trivial
light curve
\begin{equation}
f(t)=\eEdd\,L_{Edd}\,\Theta(t)\Theta(t_{Q,\,LB}-t),
\end{equation}
where $\Theta$ is the Heaviside step function, it has no strong
theoretical or observational motivation. For present purposes, models
where quasars live arbitrarily long with slowly evolving
mean volume emissivity or mean light curve 
\citep[e.g.][]{SB92,HM00,KH00} are equivalent to the ``light bulb''
scenario, as they still assume that quasars observed at a luminosity
$L$ radiate at that approximately constant luminosity over some
universal lifetime $t_{Q,\,LB}$ at a particular redshift.

We also consider a variant of the ``light bulb'' scenario, which we
term the ``Eddington limited'' model, where a black hole accretes at a
fixed Eddington ratio $\eEdd$ from an initial mass $M_{i}$ to a final
mass $M_{f}$ (or equivalently, a final luminosity
$L_{f}=\eEdd\,L_{Edd}(M_{f})$), and then shuts off.  This gives
exponential mass and luminosity growth, with the light curve
\begin{equation}
f(t)=\eEdd\,L_{Edd}(M_{i})\,e^{\eEdd\,t/t_{S}},
\end{equation}
where $f(t)=0$ for $t>\ln{M_{f}/M_{i}}$.  The time spent in 
any logarithmic luminosity bin is constant, 
\begin{equation}
dt/\dlgL = t_{S}\,(\ln(10)/\eEdd)
\end{equation}
for $L_{i}<L<L_{f}$. This is true for any exponential light
curve $f(t)\propto e^{\pm t/t_{\ast}}$, such as that
of \citet{HL98}, with only the normalization $dt/\dlgL = t_{\ast}$
changed, and thus any such model will give identical results with
correspondingly different normalizations.  While such a
light curve allows the black hole mass and luminosity to change, 
there is no theoretical
expectation that black hole growth is well-approximated by a
constant Eddington ratio.

We compare these scenarios to a physically motivated model using
quasar lifetimes derived from simulated galaxy mergers
\citep{H05a,H05b}.  The light curves from the simulations are complex,
generally having periods of rapid accretion after ``first passage'' of
the galaxies, followed by an extended quiescent period, then a
transition to a peak, highly luminous quasar phase, and then a dimming
as self-regulated mechanisms expel gas from the remnant center after
the black hole reaches a critical mass
\citep{DSH05}.  In addition, the accretion rate at any time can be
variable over small timescales $\sim{\rm Myr}$, but even with these
complexities, the statistical nature of the light curve can be
described by simple forms.  \citet{H05b} find that
the total quasar lifetime $t_{Q}(L'>L)$ above a given luminosity $L$
is well-approximated by a truncated power law, with
\begin{equation}
t_{Q}(L'>L) = t_{9}\,(L / 10^{9}\,L_{\sun})^{\alpha}, 
\end{equation}
where $t_{9}\equiv t_{Q}(L'>10^{9}\,L_{\sun})\sim10^{9}\,{\rm yr}$,
over the range $10^{9}\,L_{\sun}<L<L_{peak}$ for a given
quasar.  $L_{peak}$ is, as above, determined by the final black hole
mass, $L_{peak}\approx
L_{Edd}(M_{f})$.  Over a wide range of $L_{peak}$ (from
$\sim10^{10}-10^{14}\,L_{\sun}$), \citet{H05b} also find that $\alpha$ 
is a function of $M_{f}$ (or $L_{peak}$), given by
$\alpha=\alpha_{0}+\alpha'\,\log{L_{peak}}$, with $\alpha=-0.2$ (the
approximate slope of the Eddington-limited case) as an upper
limit. This reflects the fact that larger quasars have shallower
slopes as they spend more time at higher luminosities up to some
larger peak luminosity.  The time spent in any logarithmic luminosity
interval in this range is then simply
\begin{equation}
dt/\dlgL = |\alpha|\,t_{9}\,(L / 10^{9}\,L_{\sun})^{\alpha}. 
\end{equation}

\begin{figure}
    \centering
    %\plotone{f1.ps}
    \includegraphics[width=3.5in]{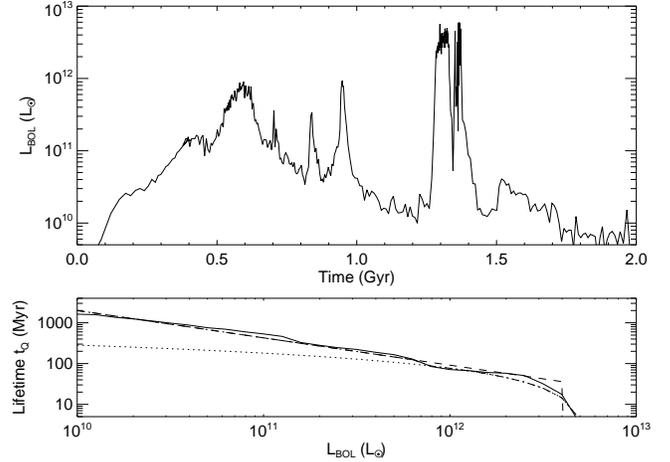}
    \caption{A typical quasar light curve (top) and 
    lifetime as a function of luminosity (bottom), for 
    a major merger of Milky Way-like galaxies with a final black hole mass
    $M_{f}=3\times10^{8}\,M_{\sun}$ \citep{H05a}. 
    The simulation lifetime (solid line) is compared 
    to the expected Eddington-limit model lifetime (dotted), the power law approximated 
    lifetime (dashed), and the mixed power law-Eddington model (dash-dot), 
    from the formulae in \S2.%~\ref{sec:lifetimes}. 
    \label{fig:lf.lightcurve}}
\end{figure}

In our picture, quasars spend far more time at low luminosities than
at their peaks.  This is a consequence of both pure $\eEdd=1$
Eddington-limited growth up to a final black hole mass as well as a
``brightening'' and ``dimming'' out of and into extended quiescent
phases of black hole growth around the peak quasar phase, during which
the Eddington ratio changes with time in the range $l \sim0.01-1$.  We
also consider a more complicated fit described in \citet{H05b}, where
quasar lifetimes follow the Eddington-limited model at the brightest
luminosities, down to some $L_{\ast}$, below which they are described
by a power law.  In all these cases, the best-fit break luminosity is
a nearly constant fraction of the peak luminosity $L_{\ast}\approx
0.14\,L_{peak}$.  These fits have the advantage of being accurate at
both low and high luminosities, down to lifetimes $\lesssim\,$Myr.  In
Figure 1, we show an example of the light curve from one of our
simulations (top) and the relation between the lifetime and
luminosity and various fits to it (bottom).

Our model contrasts sharply with both the ``light bulb'' and
``Eddington limited'' cases.  The ``light bulb'' scenario ignores the
evolution of the quasar as it accretes, and neglects the vast majority
of its life at luminosities below its final peak. The ``Eddington
limited'' model may be a reasonable approximation during the bright
quasar phase near the final peak luminosity, but it underestimates the
time spent at luminosities $L\lesssim0.1\,L_{peak}$ by as much as two
orders of magnitude \citep{H05b}.  
Furthermore, in our model, quasars can evolve significantly on 
short timescales ($\sim10$\,Myr), which means that our interpretation 
yields a statistical mix of contributions from different peak luminosities 
and Eddington ratios at any observed luminosity at a given redshift, 
which contrasts strongly with any model in which the quasar light curve 
or emissivity evolves over cosmic time.
We note that our arguments do not
in any way depend on the obscuration model described in
\citet{H05a,H05b}, as we consider only low enough redshifts that
quasar LFs should be reasonably complete and
therefore only ``intrinsic'' quasar lifetimes are relevant.
Incorporating such effects, however, does not qualitatively change our
conclusions.  Moreover, the simulations from which we derive our
lifetimes cover the complete range of intrinsic and observed
luminosities of the LFs considered here, and show
smoothly-changing properties in good agreement with the fits described
above for all simulations in this range.

\section{Results\label{sec:results}}

Given quasar lifetimes as functions of luminosity and
peak luminosity, the LF is a
convolution of the lifetime with the intrinsic distribution of
sources with a given $L_{peak}$,
\begin{equation}
\Phi(L)\propto\int{\frac{dt(L,L_{peak})}{\dlgL}\,\nLP}\,d\log(L_{peak}). 
\end{equation}
Here, $\nLP$ is the rate at which sources in a logarithmic interval
in $L_{peak}$ are created or ``activated'' per unit volume at some 
redshift, %, $\nLP\equiv dN(L'_{peak}>L_{peak})/d\log(L_{peak})$,
and $\Phi(L)$ is the observed number density of sources per
logarithmic interval in $L$.  This formulation implicitly accounts for
the ``duty cycle'' (the fraction of active quasars at a given
time), which is proportional to the lifetime at a given luminosity. As
we do not attempt to model the number of halos containing quasars and
wish to consider wide classes of lifetime models (which may have
different normalizations), we do not concern ourselves with the
absolute normalization of $\Phi(L)$ or $\nLP$.

For the ``light bulb'' and ``Eddington limited'' scenarios
respectively, this
convolution and the corresponding deconvolution are trivial, giving
\begin{equation}
\dot{n}_{LB}(L_{peak}) \propto \Phi(L=L_{peak}) 
\end{equation}
\begin{equation}
\dot{n}_{EL}(L_{peak}) \propto \frac{d\Phi}{\dlgL}\raisebox{-3pt}{\huge $\mid$}_{L=L_{peak}} 
\end{equation}
for the intrinsic $L_{peak}$ distributions. 
\citet{SW03} use a similar method to de-convolve 
the distribution of sources and consider in detail a very wide range of models for the 
quasar light curve and distribution of Eddington ratios, but none of the 
models they consider include the fundamentally important feature of 
the lifetime increasing with decreasing luminosity to the lowest observed quasar 
luminosities, and further they do not allow the Eddington ratio distribution to 
depend on final black hole mass. As a result, every example in the range of models 
they consider, as well as those in previous works, results in an $\nLP$ distribution 
qualitatively similar to the observed LF, increasing monotonically to lower luminosities 
(in the absence of an arbitrary truncation). 
However, in the case of our 
luminosity-dependent lifetimes, we must solve numerically for
$n_{LDL}$, fitting to a given LF, and find a fundamentally different result.

\begin{figure}
    \centering
    %\plotone{f2.ps}
    \includegraphics[width=3.5in]{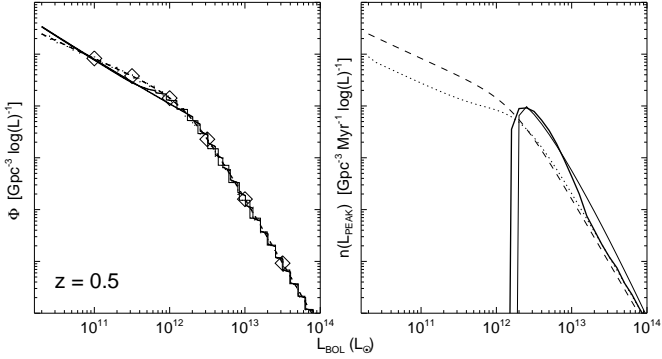}
    \caption{Deconvolved intrinsic distributions of peak luminosities $\dot{n}(L_{peak})$, 
    from fitting to the \citet{Boyle00} LF (diamonds) at redshift $z=0.5$. 
    Upper panel shows the fitted LF, lower panel the $\dot{n}(L_{peak})$
    distribution for ``light bulb'' models (dashed), ``Eddington limited'' models (dotted), 
    and luminosity dependent lifetimes fitted to power laws (thin) and 
    fitted to combined power laws and Eddington growth (thick). 
    \label{fig:lf.normal}}
\end{figure}

Figure~\ref{fig:lf.normal} shows the best-fit $\nLP$ distributions for
the above cases fitted to the \citet{Boyle00} optical quasar
LF at redshift $z=0.5$, as well as the best-fit LF.
For luminosity-dependent lifetimes, we take a
power law dependence on $L_{peak}$ from \citet{H05b}, with
$\alpha=-0.95+0.32\log(L_{peak}/10^{12}\,L_{\sun})$ for pure truncated
power-law fits and $\alpha=-0.98+0.32\log(L_{peak}/10^{12}\,L_{\sun})$
for fits with a power-law mapped onto an Eddington-limited fit at
near-peak luminosities.  For ease of comparison with black hole mass
and Eddington ratios, we rescale all B-band quantities to the
bolometric luminosity using the corrections of
\citet{Marconi04}.  In all cases, it is possible to reproduce the
double-power law LF and ``break'' luminosity
$L_{brk}$ quite accurately, and the $\nLP$ distributions are similar
at luminosities above $L_{brk}$. However, below $L_{brk}$, a proper
accounting of the luminosity dependence of quasar lifetimes results in
a radically different $\nLP$ distribution, with the shallow slope of
the LF a direct reflection of the slope in the
lifetime vs.\ luminosity relation, dominated by quasars at the peak of
the $\nLP$ distribution, which then
determines the ``break'' luminosity of the observed LF,
with $L(n_{LDL,\,max})\sim L_{brk}$.
Note that the steepness of the $\nLP$ cutoff is an artifact of extrapolating the LF to 
luminosities well below those observed -- any strong decline below $L_{brk}$ 
gives an equivalent fit to the actual data. 

It is possible that objects which have already undergone a quasar
phase may re-activate, or that black holes may already be very large
when AGN activity first begins. To determine whether or not this
changes our results, we consider the most extreme possible case, in
which we return to the simulations of \citet{H05a,H05b} and re-fit the
quasar lifetimes but ignore all black hole activity prior to the final
merger. This neglects all early accretion activity as black holes
build up to a significant mass, as well as weaker, sub-Eddington
accretion during the merger as gravitational torques from early merger
stages funnel gas to the galaxy centers. Therefore, we expect this case
to set a strong lower limit to the lifetime at low luminosities,
regardless of the mechanism driving quasar activity. However, we still
find that these lifetimes are well above the Eddington-limited model
expectations, albeit with shallower best fit power law slopes
$\alpha=-0.77+0.30\log(L_{peak}/10^{12}\,L_{\sun})$ and a slightly
lower (constant) normalization.

It has also been argued from observations of stellar
black hole binaries that a transition between
accretion states occurs at a critical Eddington ratio $\mdot\equiv
M/M_{Edd}$, from radiatively inefficient accretion flows at low
accretion rates \citep[e.g.,][]{EMN97} to radiatively efficient
accretion through a standard \citet{SS73} disk. Although the critical
Eddington ratio for supermassive black holes is uncertain,
observations of black hole binaries \citep{Maccarone03} as well as
theoretical extensions of accretion models \citep[e.g.,][]{MLMH00}
suggest $\mdot_{crit}\sim0.01$.  We examine whether this can
change our conclusions by re-fitting quasar lifetimes from our
simulations. Because we assumed
a constant radiative efficiency
$L=\epsilon_{r}\,\dot{M}\,c^{2}$ with $\epsilon_{r}=0.1$, we account
for this effect by multiplying the simulation luminosity at all times
by an additional ``efficiency factor'' $f_{eff}$ which depends on the
Eddington ratio $\eEdd=L/L_{Edd}$ (given an a priori constant assumed
efficiency, $\eEdd=\mdot$ always),
\begin{equation}
  f_{eff} = \left\{ \begin{array}{ll}
      1  & \mathrm{ if\ } \eEdd > 0.01 \\
      100\,\eEdd & \mathrm{ if\ } \eEdd \leq 0.01.
\end{array}
    \right.
\end{equation}
This choice for the efficiency factor follows from ADAF models
\citep{NY95} and ensures that the radiative efficiency is continuous
at the critical Eddington ratio $\eEdd_{crit}=0.01$. We expect this to
set an extreme limit to the impact of this transition, as we do not
incorporate this effect dynamically in the simulations, which would
slow ``blowout'' after the peak quasar phase and increase lifetimes at
low luminosities.  We again find that power-law lifetimes
at low luminosities are much larger than the expectation from the
Eddington-limited model, with best-fit slopes
$\alpha=-0.94+0.33\log(L_{peak}/10^{12}\,L_{\sun})$.

\begin{figure}
    \centering
    %\plotone{f3.ps}
    \includegraphics[width=3.5in]{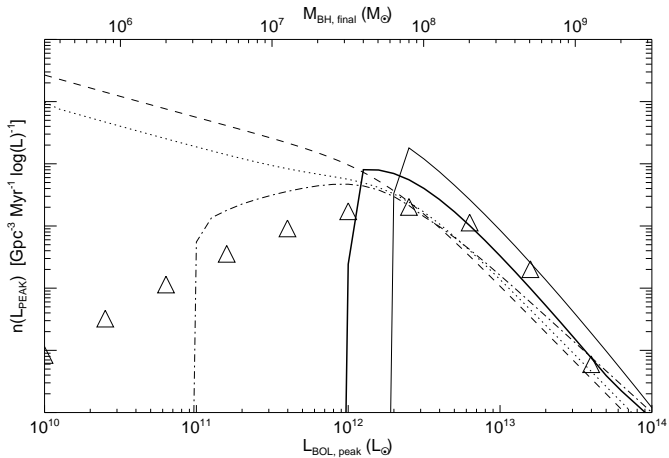}
    \caption{As Figure~\ref{fig:lf.normal}, lower panel, but with lifetimes 
    recalculated considering only times after the onset of peak quasar activity (thick), 
    or decreasing the radiative efficiency below a critical Eddington ratio (thin). The 
    dot-dashed line uses lifetimes calculated including both effects.     
    ``Light bulb'' (dashed) and ``Eddington limited'' (dotted) models are shown for comparison.
    Triangles show the distribution of black hole masses expected from the 
    distribution of early-type velocity dispersions \citep{Sheth03} and the 
    $M_{BH}-\sigma$ relation from \citet{Tremaine02}.
    \label{fig:lf.limits}}
\end{figure}

Figure~\ref{fig:lf.limits} shows the best-fit $\nLP$ distributions and
LFs for these cases, in the manner of
Figure~\ref{fig:lf.normal}. We also plot as an extreme limit the
results obtained applying both the radiative efficiency correction
above and considering only times during and after the final
merger. This case gives the shallowest slopes,
$\alpha=-0.50+0.5\log(L_{peak}/10^{12}\,L_{\sun})$.  Although the
shallower slopes broaden the $\nLP$ distribution, our results are
qualitatively unchanged.  We have further considered extreme cases
such as making our original slopes shallower by $3\sigma$, a factor
$\sim2$, and find nearly identical results.

\section{Conclusions\label{sec:conclusions}}

We consider realistic quasar light curves and lifetimes derived from 
hydrodynamical simulations of quasar activity, and find that this results 
in a qualitatively different form and evolution of the distribution in peak quasar luminosities $\nLP$
(equivalently, final black hole masses or host system properties) than that derived or implied in previous 
works \citep[e.g.,][]{SB92,HL98,HM00,KH00,WL03,V03,HQB04}. 
We describe the resulting new interpretation of the quasar luminosity function
(LF) in which the steep bright-end consists of quasars
radiating near their Eddington limits and is directly related to
the distribution of intrinsic peak luminosities (or final black hole
masses) as has been assumed previously.  However, the shallow,
faint-end of the LF describes black holes either growing efficiently
in early stages of activity or in extended, quiescent states going
into or coming out of a peak bright quasar phase, with Eddington
ratios generally between $l \sim0.01$ and 1. These are not objects which 
were active at some earlier cosmic time and have since ``shut down'', 
as are invoked in some ``light bulb'' models, but correspond to the evolution 
of quasars in mergers over short timescales and critically related (both 
physically and statistically) to the observed 
bright, Eddington-limited quasar phase. The ``break'' luminosity
in the LF corresponds directly to the {\em peak} in the distribution
of intrinsic quasar properties $\nLP$.  As such, observations of this
break and its evolution present a powerful new probe of the quasar
population.

We find that the observed quasar LF can be reproduced to high accuracy
with a model of quasar lifetimes which depends on both the observed
luminosity and peak luminosity of a quasar, for some
distribution of peak luminosities $\nLP$.  We demonstrate this for the
\citet{Boyle00} LF at $z=0.5$, but find
identical qualitative results for the \citet{Miyaji00} and
\citet{Ueda03} soft and hard X-ray LFs, over a wide range of redshifts
$z=0-3$. This is not trivial, as it is in the case of simpler
models of quasar lifetimes (in which the $\nLP$ distribution can be
directly recovered from the LF).  If the faint-end slope of the LF is
too shallow, power-law luminosity dependent lifetimes will be too
steep to reproduce it.  If the faint-end slope is too steep, our
lifetime model can reproduce the LF, but without a characteristic
qualitative difference in behavior from ``light bulb'' or
Eddington-limited models.  In fact, it is a strong argument in
favor of our picture that observed faint-end slopes lie just within
the expected range, for any peaked population $\nLP$.

Our new interpretation of the quasar LF immediately
explains several other observations.  The distribution of velocity
dispersions in early-type systems (expected to have undergone
merger-driven AGN activity) from \citet{Sheth03} turns over and
decreases below $\sigma\approx160\,{\rm km s^{-1}}$, which from the
$M_{BH}-\sigma$ relation implies that the black hole mass distribution 
for remants of major mergers of massive galaxies 
should turn over and decrease below $M_{BH}\sim10^{8}\,M_{\sun}$
($L_{peak}\sim10^{12}\,L_{\sun}$).  Moreover, \citet{HQB04} find that
black holes with masses less than $\sim10^{7}\,M_{\sun}$ must be rare at high 
redshift to avoid significantly overpredicting the counts of bright radio
sources. These observations are exactly as predicted
by our deconvolution of the LF, where at high redshift $\nLP$ is shifted 
to higher peak luminosity (black hole mass) and thus fewer low-mass black holes are 
produced.  The $\nLP$
distribution based on traditional models of quasar lifetimes,
however, would necessarily predict that the vast majority of
quasars, especially at high redshift where the number of quasars is larger, 
have final black hole masses well
below this limit, with the number density increasing towards lower
black hole masses. The shift of $\nLP$ with time in our model 
can explain the entire observed range of the $z=0$ black hole mass function
\citep[e.g.,][]{Marconi04}.

The distribution of observed Eddington ratios is also a natural
consequence of our model, and does not invoke a particular probability
distribution across sources to match observations.  Quasars observed
around and below the ``break'' in the LF (which
dominate the total population) may be accreting at the Eddington rate,
as they build up early in their lives, or may be in quiescent phases
going into or out of their peak quasar activity. For these quiescent
phases, both comparison of the lifetime power laws in
\S3\ %~\ref{sec:results}
and direct calculation of the lifetime as a
function of Eddington ratio \citep{H05a,H05b} show that they are
dominated by relatively large Eddington ratios $l \gtrsim0.1$.  Below
$\eEdd\sim0.01$, quasars at the peak of the $\nLP$ distribution will
be below limits measured in most surveys.  Therefore, we expect a
distribution of Eddington ratios concentrated between $l \sim0.1-1$,
with a small tail down to $l \sim0.01$ and mean Eddington ratios
increasing with luminosity, exactly as observed \citep{Vestergaard04}.

The sharp contrast between the intrinsic distribution of peak quasar
luminosities and the observed LF provides clean and elegant tests of
our theory.  In our interpretation, the bright and faint ends of the
LF correspond statistically to similar mixes of galaxies, but in
various stages of evolution.  However, in all other competing
scenarios, the quasar luminosity is directly related to the mass of
the host galaxy.  Therefore, any observational probe that
differentiates quasars based on their host galaxy properties such as,
for example, the dependence of clustering of quasars on luminosity,
can be used to discriminate our picture from older models. Early evidence 
from both quasar-quasar \citep{Croom05} and 
quasar-galaxy \citep{Adelberger05} correlations support this picture 
and suggest that quasar hosts have a well-defined characteristic mass, 
exactly as predicted by our model with a peaked $\nLP$ and in stark 
contrast to previous models in which $\nLP$ continues to increase below 
the break in the luminosity function. 

Our modeling of realistic quasar lifetimes and the 
resulting new $\nLP$ distribution and interpretation of the quasar luminosity 
function have a wide range of implications, . Theoretical predictions of the 
active and relic supermassive black hole density and mass functions, the 
contribution of quasars to the X-ray and infrared background, the observed 
Eddington ratio distribution as a function of luminosity and redshift, the 
evolution of characteristic black hole masses and quasar host galaxy masses 
with redshift, quasar correlation functions and bias, the relation between quasar 
and starburst or luminous infrared galaxy luminosity functions, the role of 
quasars in reionization, the population of high-redshift radio sources, 
means to discriminate between pure density and pure luminosity evolution of 
quasar populations at very high redshift, and the distribution and evolution of 
quasar host properies all depend on this quantity and must be revised in 
this new model. Semi-analytical models and cosmological simulations 
have focused on reproducing $\nLP$ distributions qualitatively similar to the 
observed LF based on idealized models of the quasar lifetime, but should  
instead attempt to reproduce the fundamentally different $\nLP$ distribution implied 
by our realistic, physically motived models of the quasar light curve and 
luminosity-dependent quasar lifetimes. 

\acknowledgments
This work was supported in part by NSF grants ACI
96-19019, AST 00-71019, AST 02-06299, and AST 03-07690, and NASA ATP
grants NAG5-12140, NAG5-13292, and NAG5-13381.
The simulations
were performed at the Center for Parallel Astrophysical Computing at the 
Harvard-Smithsonian Center for Astrophysics.

\end{document}